\documentclass[final,3p,times,twocolumn]{elsarticle}

\usepackage{graphicx}

\usepackage{amssymb}
\usepackage{amsmath}
\usepackage{amsthm}
\usepackage{bm}

\journal{Journal ofElectronSpectroscopy and Related Phenomena}

\begin{document}

\begin{frontmatter}



\title{Calculation of angle-resolved photo emission spectra within the one-step
model of photo emission - recent developments}


\author{J. Min\'ar, J. Braun, S. Mankovsky and H. Ebert}

\address{University Munich, Dept. Chemistry, Munich, Germany}

\begin{abstract}
Various technical developments enlarged the potential of angle-resolved photo emission (ARPES) tremendously
during the last one or two decades. In particular improved momentum and energy resolution as well as the use
of photon energies from few eV up to several keV makes ARPES a rather unique tool to investigate the electronic
properties of solids and surfaces. Obviously, this rises the need for a corresponding theoretical formalism that allows
to accompany experimental ARPES studies in an adequate way. As will be demonstrated by several examples this goal
could be achieved by various recent developments on the basis of the
one-step model of photo emission: The spin-orbit
induced Rashba-splitting of Shockley-type surface states is discussed using a fully relativistic description. The
impact of chemical disorder within surface layers can be handled by means of the Coherent Potential Approximation
(CPA) alloy theory. Calculating phonon properties together with the corresponding electron-phonon self-energy allows
a direct comparison with features in the ARPES spectra caused by electron-phonon interaction. The same holds for
the influence of electronic correlation effects. These are accounted for by means of the dynamical mean field theory
(DMFT) that removes the most serious short comings of standard calculations based on the standard local density
approximation (LDA). The combination of this approach with the CPA allows the investigation of correlated transition
metal alloys. Finally, accounting for the photon momentum and going beyond the single scatter approximation for the
final state allows to deal quantitatively with ARPES in the high-energy regime (HAXPES) that reduces the influence of
the surface on the spectra and probing primarily the bulk electronic structure this way. Corresponding calculations of
ARPES spectra, however, have to deal with thermal vibrations in an adequate way. For this, a new scheme is suggested
that makes use of the CPA.
\end{abstract}


\end{frontmatter}


\section{Introduction}
\label{intro}
The spectrum of one-particle excitations of a system of correlated electrons in a solid is a fundamental question in
condensed-matter physics. The theoretical understanding of the excitation spectrum poses a long-standing and not yet
generally solved problem. Within the independent-electron approximation the spectrum is simply given in terms of the
one-particle eigenenergies of the Hamiltonian. Analogously, it is widely accepted to interpret a measured photo emission
spectrum by referring to the results of band-structure calculations that are based on density functional theory (DFT)
and the local density approximation (LSDA) \cite{HK64,KS65,JG89}. Such an interpretation is questionable since there is
actually no direct correspondence between the Kohn-Sham eigenenergies and the one-particle excitations of the system
\cite{JG89,Bor85}. For an in principle correct description of the excitation energies, the local LSDA exchange-correlation
potential has to be supplemented by the non-local, complex and energy-dependent self-energy which leads to the Dyson
equation \cite{JG89,SK04} instead of the Schr\"odinger- or Dirac-type equation in the non- or fully relativistic
Kohn-Sham scheme.

To overcome these deficiencies we have developed a generalized approach by accounting properly for electronic
correlations beyond the LSDA. For this purpose a general non-local, site-diagonal, complex and
energy-dependent self-energy $\Sigma^{DMFT}$ \cite{GKKR96,KSH+06,Hel07}, has been included self-consistently in the fully relativistic
Korringa-Kohn-Rostoker multiple scattering theory (SPRKKR) \cite{MCP+05}. Within this scheme the self-energy is determined
by a self-consistent dynamical mean field theory (DMFT) calculation. This method is straightforwardly applicable for
electronic structure calculations of 3D- and 2D-systems with perfect lateral translational invariance and arbitrary number
of atoms per unit cell. Furthermore, the calculational scheme has been extended to the case of disordered alloys
\cite{MCP+05}, to include many-body correlation effects in the electronic structure and photo emission calculations of
these class of complex compound materials.

Provided that the self-energy is known, one can deduce a PES/IPE ``raw spectrum'' from the solution of the Dyson equation.
To achieve a reliable interpretation of experiments, however, it is inevitable to deal with so-called matrix-element
effects which considerably modify and distort the raw spectrum. Above all, the wave-vector and energy dependence of the
transition-matrix elements has to be accounted for. These dependencies are known to be important and actually cannot be
neglected. They result from strong multiple-scattering processes which dominate the electron dynamics in the low-energy
regime of typically 1-200~eV \cite{Pen74}. The transition-matrix elements also include the effects of selection rules which
are not accounted for in the raw spectrum. Strictly speaking, it can be stated that the main task of a theory of
photo emission is to close the gap between the raw spectrum obtained by LSDA or LSDA+DMFT electronic-structure calculations
and the experiment. The most successful theoretical approach is the so-called one-step model of photo emission as originally
proposed by Pendry and co-workers \cite{Pen74,Pen76,HPT80}. In the following a short overwiew will be given on the recent
extensions of the one-step model which are connected with correlation effects, disordered alloys, phonon related features
and high excitation energies far beyond the ultraviolet regime. 

\section{Recent developments in the one step model of photo emission}
\label{recent}
The main idea of the one-step model is to describe the actual excitation process, the transport of the photoelectron to
the crystal surface as well as the escape into the vacuum \cite{BS64} as a single quantum-mechanically coherent process
including all multiple-scattering events. Within this model self-energy corrections, which give rise to damping in the
quasi-particle spectrum, are properly included in both the initial and the final states. This for example allows for
transitions into evanescent band gap states decaying exponentially into the solid. Similarly the assumption of a finite
lifetime for the initial states gives the opportunity to calculate photo emission intensities from surface states and
resonances. Treating the initial and final states within the fully relativistic version of layer Korringa-Kohn-Rostoker
(KKR) theory \cite{Ebe00}, it is a straight forward task to describe complex layered structures like thin films and
multilayers within the photo emission theory. Furthermore, the surface described by a barrier potential can be easily
included into the multiple-scattering formalism as an additional layer. A realistic surface barrier model which shows
the correct asymptotic behavior has been introduced, for example, by Rundgren and Malmstr\"om \cite{MR80}.

We start our considerations by a discussion of Pendry's formula for the photocurrent which defines the one-step model of
PES \cite{Pen76}:
\begin{equation} I^{\rm PES} \propto {\rm Im}~ \langle E_f,
{\bf k}_{\|} | G_{2}^+ \Delta G_{1}^+ \Delta^\dagger G^-_{2} |
E_f, {\bf k}_{||} \rangle \: .
\label{eq:pendry}
\end{equation}
The expression can be derived from Fermi's golden rule for the transition probability per unit time \cite{Bor85}, with
$I^{\rm PES}$ denoting the elastic part of the photocurrent. Vertex renormalizations are neglected. This excludes inelastic
energy losses and corresponding quantum-mechanical interference terms \cite{Pen76,Bor85,CLR+73}. Furthermore, the
interaction of the outgoing photoelectron with the rest system is not taken into account. This ``sudden approximation'' is
expected to be justified for not too small photon energies. We consider an energy-, angle- and spin-resolved photo emission
experiment. The state of the photoelectron at the detector is written as $|E_f, {\bf k}_{\|} \rangle$, where
${\bf k}_{\|}$ is the component of the wave vector parallel to the surface, and $E_f$ is the kinetic energy of the
photoelectron. The spin character of the photoelectron is implicitly included in $|E_f, {\bf k}_{\|} \rangle$ which is
understood as a four-component Dirac spinor. The advanced Green function $G_{2}^-$ in Eq.\ (\ref{eq:pendry}) characterizes
the scattering properties of the material at the final-state energy $E_2 \equiv E_f$. Via
$|\Psi_f \rangle = G^-_{2}|E_f, {\bf k}_{\|}\rangle$ all multiple-scattering corrections are formally included. For an
appropriate description of the photo emission process we have to model the correct asymptotic behavior of $\Psi_f({\bf r})$
beyond the crystal surface, introducing a single outgoing plane wave characterized by $E_f$ and ${\bf k}_\|$. Furthermore,
the damping of the final state due to the imaginary part of the inner potential $iV_{0{\rm i}}(E_2)$ must be taken into
account. We thus construct the final state within spin-polarized low-energy electron diffraction (SPLEED) theory considering
a single plane wave $|E_f,{\bf k}_\|\rangle$ advancing onto the crystal surface. Using the standard layer-KKR method
\cite{Ebe00} generalized for the relativistic case \cite{Bra96,Bra01}, we first obtain the SPLEED state $-T \Psi_f({\bf r})$.
The final state is then given as the time-reversed SPLEED state ($T=-i \sigma_y K$ is the relativistic time inversion operator).
Many-body effects are included phenomenologically in the SPLEED calculation, by using a parametrized, weakly energy-dependent
and complex inner potential $V_0(E_2)=V_{0{\rm r}} (E_2)+iV_{0{\rm i}}(E_2)$ \cite{Pen74}. This generalized inner
potential accounts for inelastic corrections to the elastic photocurrent \cite{Bor85} as well as the actual (real)
inner potential, which serves as a reference energy inside the solid with respect to the vacuum level \cite{HPM+95}. Due to
the finite imaginary part $V_{0{\rm i}}(E_2)$, the flux of elastically scattered electrons is continuously reduced, and
thus the amplitude of the high-energy wave field $\Psi_f({\bf r})$ can be neglected beyond a certain distance from the
surface. 

The practical calculation starts with the Dirac Hamiltonian $h_{\rm LSDA}$ which one has to consider in the framework of
 DFT \cite{RC73,RR83}. In atomic units ($\hbar = m = e = 1,
 c=137.036$) one has:
\begin{equation} h_{\rm LSDA}({\bf r}) = - i c \mbox{\boldmath$\alpha$}
\mbox{\boldmath$\nabla$} + \beta c^{2} - c^2 + \overline{V}_{\rm LSDA}(r) + \beta
\mbox{\boldmath $\sigma$} {\bf B}_{\rm LSDA} (r)~.
\label{eq:ldaham}
\end{equation}
$\overline{V}_{\rm LSDA}(r)$ denotes the (effective) spin-independent potential, and ${\bf B}_{\rm LSDA} (r)$ is the
(effective) magnetic field given by \cite{SESG89}:
\begin{equation} \overline{V}_{\rm LSDA}(r) = \frac{1}{2} (V^{\uparrow}_{\rm LSDA}
(r)~+~ V^{\downarrow}_{\rm LSDA} (r)),
\end{equation}
\begin{equation}
{\bf B}_{\rm LSDA}(r) = \frac{1}{2} (V^{\uparrow}_{\rm
LSDA} (r)~-~ V^{\downarrow}_{\rm LSDA} (r)) \: {\bf e}_B~.
\end{equation}
The constant unit vector ${\bf e}_B$ determines the spatial direction of the (uniform) magnetization as well as the spin
quantization axis. $\beta$ denotes the usual $4 \times 4$ Dirac matrix with the nonzero diagonal elements
${\beta}_{11}={\beta}_{22}=1$ and ${\beta}_{33}={\beta}_{44}=-1$, and the vector $\mbox{\boldmath $\alpha$}$ is given by
its components $\alpha_k = \sigma_x \otimes \sigma_k$ ($k=x,y,z$) in terms of the $2 \times 2$ Pauli-matrices $\sigma_k$.

The ``low-energy'' propagator $G_{1}^+$ in Eq.\ (\ref{eq:pendry}), i.~e.\ the one-electron retarded Green function for the
initial state in the operator representation, yields the ``raw spectrum''. It is directly related to the ``bare''
photocurrent and thereby represents the central physical quantity within the one-step model. $G_{1}^+ \equiv G_{1}^+(E_i)$
is to be evaluated at the initial-state energy $E_i \equiv E_f
-\omega$, where $\omega$ is the photon energy. In the
relativistic case $G_{1}^+$ is described by a $4 \times 4$ Green matrix which has to be obtained for a semi-infinite stack
of layers. This quantity is defined by the following equation:
\begin{eqnarray}
\hspace*{-0.4cm}
\left[ E_i~-~h_{\rm LSDA}({\bf r}) \right]
G_{1}^{+}({\bf r},{\bf r}',E_{i})=\delta({\bf r}-{\bf r}')~.
  \label{eq:eomgr}
\end{eqnarray}
In order to account for strong electronic correlations beyond the LSDA-scheme one has to introduce a non-local, energy and
spin-dependent potential V. This quantity can be defined in the following way:
\begin{eqnarray}
\hspace*{-0.4cm}
V({\bf r},{\bf r}',E) &=& \delta ({\bf r}-{\bf
r}')~(\overline{V}_{\rm LSDA}({\bf r})~+~ \beta \mbox{\boldmath $\sigma$} {\bf
B}_{\rm LSDA}({\bf r})) \nonumber \\ &+& \Sigma^{(V)}({\bf r},{\bf r}',E)~+~ \beta
\mbox{\boldmath $\sigma$}{\bf \Sigma^{(B)}}({\bf r},{\bf r}',E)~,
\end{eqnarray}
with
\begin{eqnarray} \Sigma^{(V)}({\bf r},{\bf r}',E)=\frac{1}{2}({\Sigma}^{\uparrow}
({\bf r},{\bf r}',E)+{\Sigma}^{\downarrow}({\bf r},{\bf r}',E))
\end{eqnarray}
and
\begin{eqnarray}
{\bf \Sigma^{(B)}}({\bf r},{\bf r}',E)=\frac{1}{2}({\Sigma}^{\uparrow}({\bf r},{\bf
r}',E)-{\Sigma}^{\downarrow} ({\bf r},{\bf r}',E)) \: {\bf e}_B~.
\end{eqnarray}
The resulting integro-differential equation for the initial-state one-electron retarded Green function takes the form:
\begin{eqnarray}
&&
\left[E+ic\mbox{\boldmath$\alpha$}\mbox{\boldmath$\nabla$} -
\beta c^{2}+c^2)\right] G_1^{+}({\bf r},{\bf r}',E) \nonumber \\ &+&
\hspace*{-0.5cm}
\int V({\bf r},{\bf r}'',E)~G_1^{+}({\bf r''},{\bf r}',E) d {\bf r}''=
\delta({\bf r}-{\bf r}')~.
\label{eq:dirac}
\end{eqnarray}
According to the LSDA+DMFT approach realized in the framework of the fully relativistic Korringa-Kohn-Rostoker multiple
scattering theory (SPR-KKR) \cite{MCP+05} we use a self-energy ${\Sigma}^{DMFT}_{\alpha}(E)$ calculated self-consistently using
dynamical mean-field theory (DMFT) \cite{GKKR96,KSH+06,Hel07}.

The explicit form in relativistic notation is given by:
\begin{eqnarray}
V_{\begin{substack} {\Lambda \Lambda'}
   \end{substack}} ({\bf r},{\bf r}',E) &=& (\overline{V}({\bf r}) +
\hat{\mbox{\boldmath${\sigma}$}} B({\bf r}))~\delta({\bf r-r'})
\nonumber \\ &+& \Sigma^{DMFT}_{\begin{substack} {\Lambda \Lambda'}
               \end{substack}} (E)~\delta_{l2} \delta_{l'2}~.
\end{eqnarray}
The spin-orbit quantum number  $\kappa$ and magnetic quantum number $\mu$ were combined in the symbol
$\Lambda=(\kappa, \mu)$. \\

For disordered alloys the coherent potential approximation (CPA) scheme can straightforwardly be extended to include
the many-body correlation effects for disordered alloys \cite{MCP+05}. According to the LSDA+DMFT approach realized in
the framework of the fully relativistic SPR-KKR multiple scattering theory we use a self-energy $\Sigma^{DMFT}_{\alpha}(E)$
calculated self-consistently using dynamical mean field theory \cite{GKKR96,KSH+06,Hel07} for the component $\alpha$ of an alloy.
Within the KKR approach the local multi-orbital and energy dependent self-energy $\Sigma^{DMFT}_\alpha (E)$ is
directly included into the single-site matrix $t_\alpha$ for the component $\alpha$, respectively by solving the
corresponding Dirac Eq.~(\ref{eq:dirac}). Consequently, all the relevant physical quantities connected with the Green's
function and used for photo emission calculations, as for example, the matrix elements or scattering matrices like
$t^{CPA}$ contain the electronic correlations beyond the LSDA scheme. A detailed description of the generalizated
one-step model for disordered magnetic alloys can be found in \cite{BMM+10}. Here we present the expression for the
CPA photocurrent only. Following Durham \emph{et al.} \cite{Dur81,GDG89} we have to perform the configurational average
for the various contributions to the photocurrent:
\begin{eqnarray}
\hspace*{-0.4cm}
<I^{\rm PES} (\epsilon_f, {\bf k}_{\|})> &=& <I^{\rm
a} (\epsilon_f, {\bf k}_{\|})> + <I^{\rm m} (\epsilon_f, {\bf
k}_{\|})> \nonumber \\ &+& <I^{\rm s} (\epsilon_f, {\bf k}_{\|})> +
<I^{\rm inc} (\epsilon_f, {\bf k}_{\|})> \nonumber \\ &&
\end{eqnarray}
When dealing with disorder in the alloys, an additional contribution $I^{\rm inc}$, the so called ''incoherent'' term
appears. This contribution to the alloy photocurrent appears because the spectral function of an disordered alloy
\cite{FS80} is a non single-site quantity. In fact this contribution is closely connected with the presence
of the irregular wave functions well-known from the spherical representation of the Green function G$^{+}_1$. 

Within the next section we present some selected results from photo emission investigations on various systems to
demonstrate the great flexibility of the method described above. We start with a photo emission study on MgO/Fe(001)
performed for photon energies in the hard X-ray regime.

\section{High energy photo emission spectroscopy}
Recently, the one-step model has been extended to the soft X-ray regime \cite{VMB+08,PMS+08}. Starting with a new ansatz
for the final state which generalizes the so called single-scatter approach, it becomes possible to account for the lattice
for the complete photo emission calculation. This method is applicable also for higher photon energies ranging from 1 keV
up to more or less than 10 keV. Increasing the photon energy means increasing the bulk sensitivity of the corresponding
photo emission data. We demonstrate this effect by comparing spectra obtained by photoelectron excitation with 1 keV and
6 keV radiation from the clean Fe(001) surface and from the overlayer system MgO/Fe(100) with 8 ML MgO on Fe. Fig.~1a presents
the spectrum for the clean Fe(001) surface and a photon energy of 1 keV. As expected for this photon energy regime, the bulk
sensitivity is enhanced and the surface emission is reduced to a negligible extent. Also the relative intensity fraction
of the sp-bands is obviously increased. Going to a much higher photon energy of 6 keV the d-band intensity is strongly
reduced when compared with sp-band related features. This is shown in Fig.~1b. In a next step we put 8 ML of MgO on the
Fe(001) surface and repeat our calculations for both photon energies. This is shown in the lower panel of Fig.~1. At 1 keV
mostly the MgO bulk band-structure is visible in Fig.~1c. This is due to the thickness of the MgO film which consists of
8 ML. Increasing the photon energy to 6 keV we expect due to the much larger mean free path length of the photo electron
that Fe-related features will recover, namely the Fe sp-states. This is clearly demonstrated by Fig.~1d which represents
the corresponding photo emission spectrum calculated for the 8ML MgO/Fe(001) system at a photon energy of 6 keV. Besides the
effect that more than one Brillouin zone is visible for a fixed escape angle regime due to the very high photon energy it
is undoubtly observable that nearly all MgO related features are vanished in the intensity plot. This, for example, should
give access to buried interfaces.
 
\label{haxpes}
\label{figure1}
\begin{figure}[t]
\includegraphics[width=0.23\textwidth,clip]{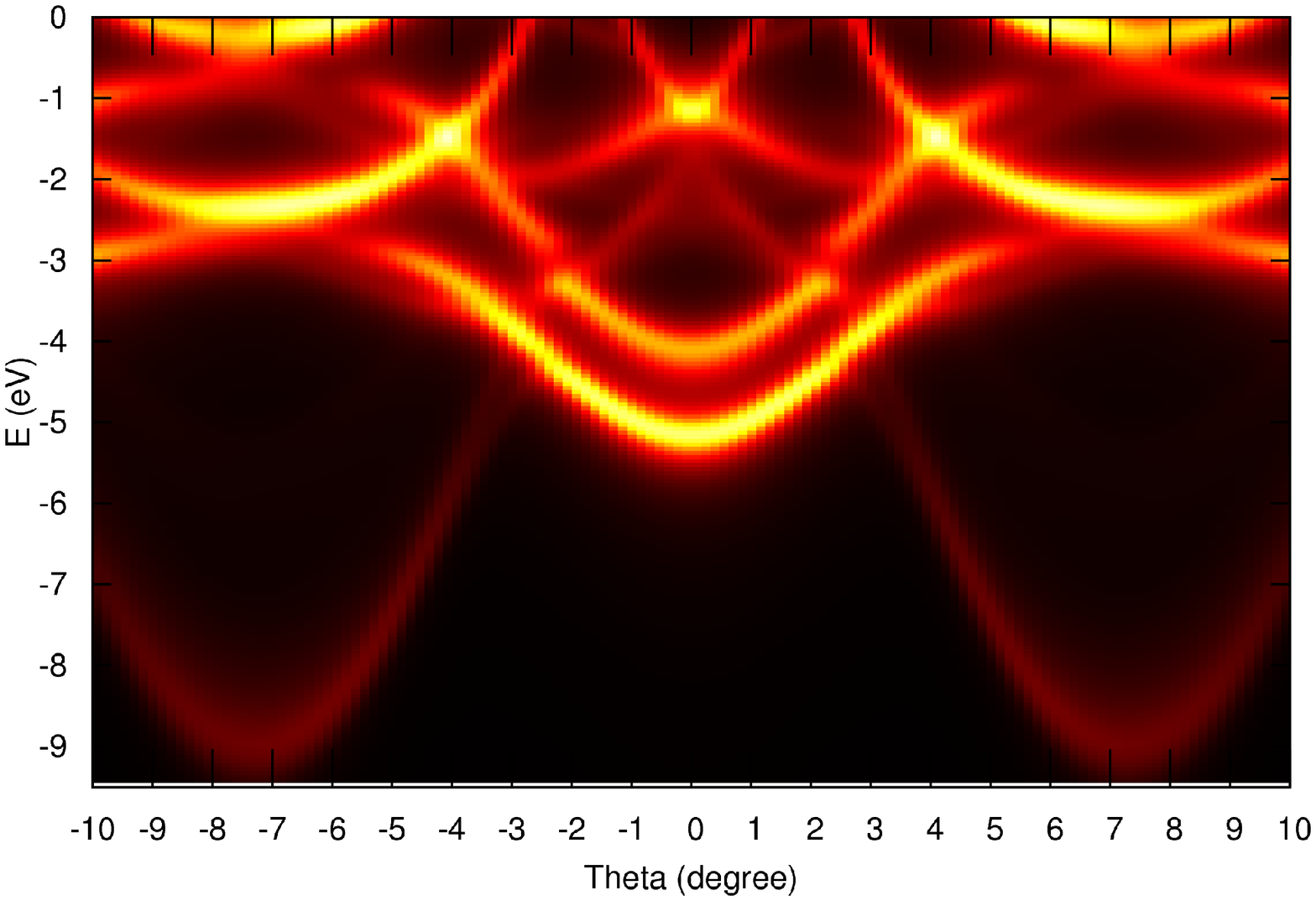}
\includegraphics[width=0.23\textwidth,clip]{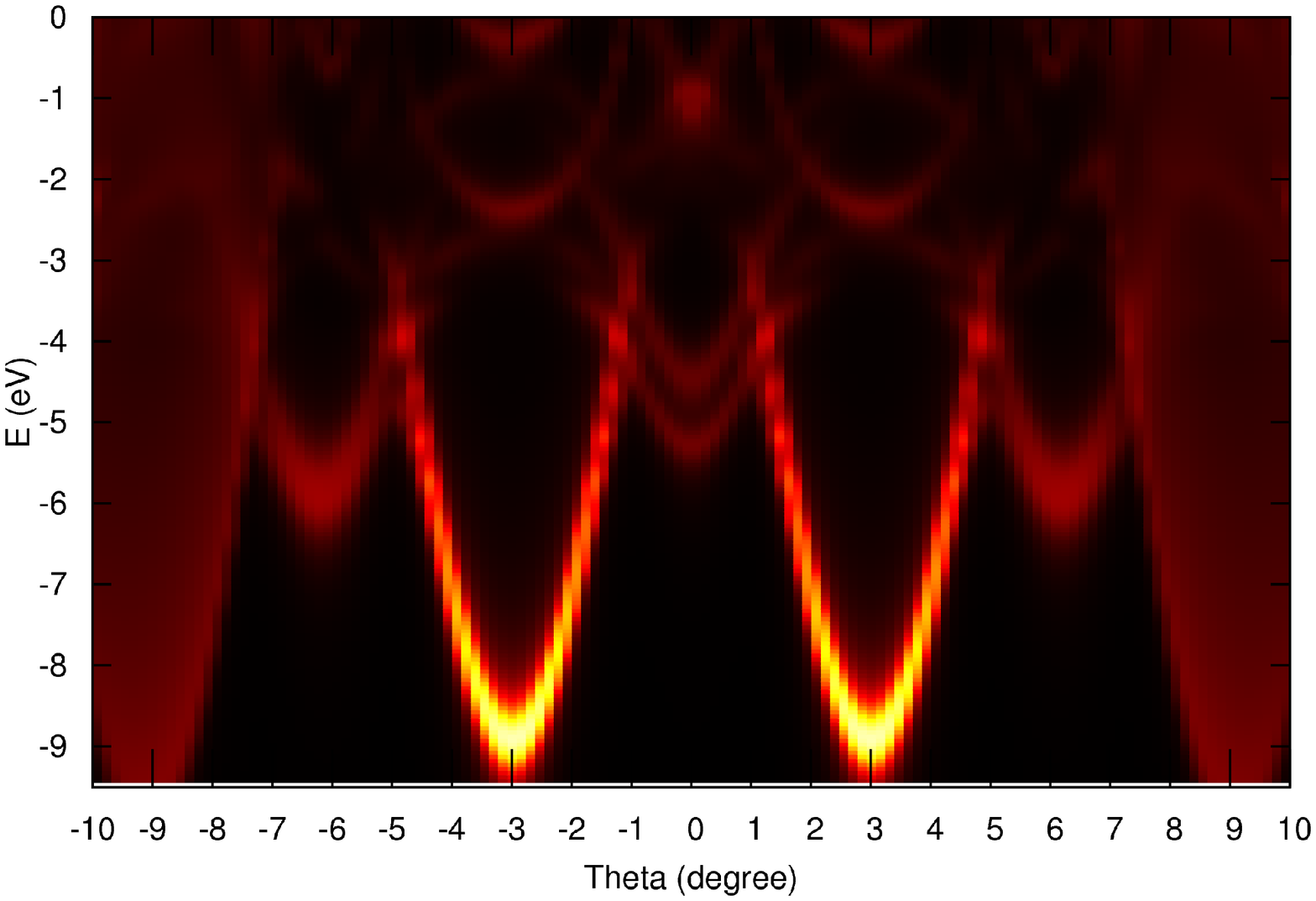}\\
\includegraphics[width=0.23\textwidth,clip]{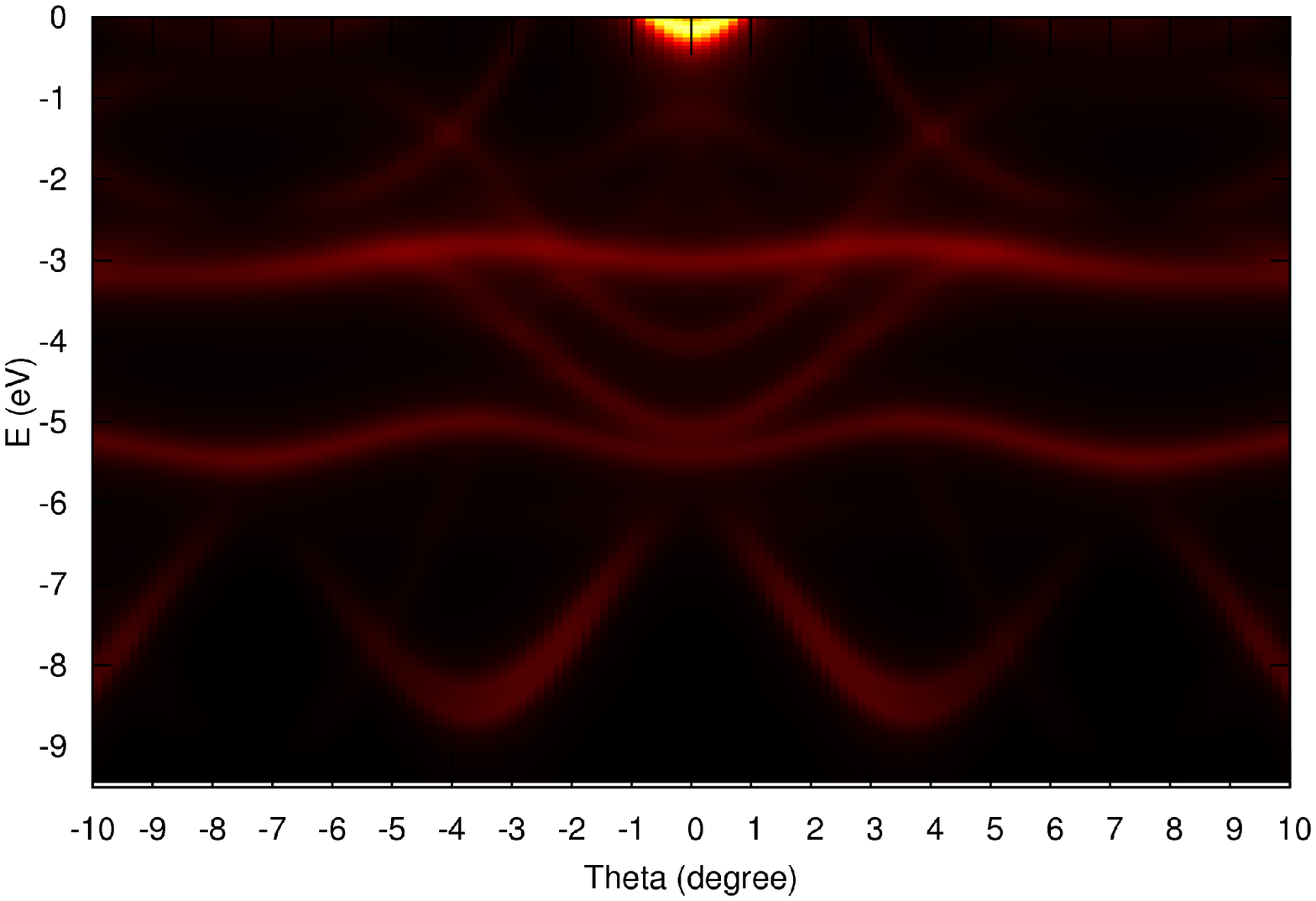}
\includegraphics[width=0.23\textwidth,clip]{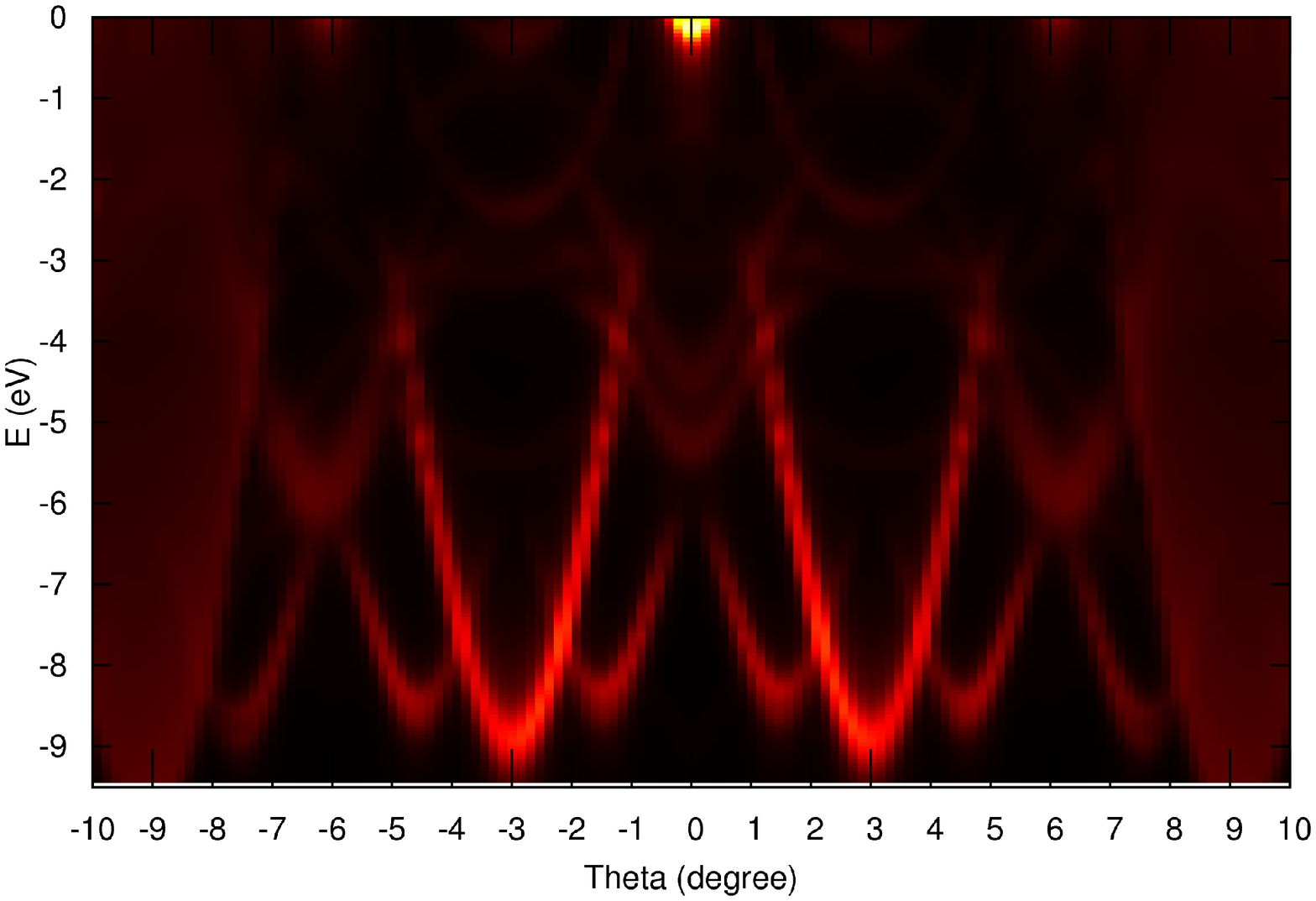}
\caption {(Color online) Photoemission spectra calculated for clean Fe(001) (upper panel, Fig.~1a and 1b). Left side
shows the intensity distribution obtained for 1 keV, right side represents the corresponding spectrum at 6 keV. Lower
panel shows theoretical spectra for the overlayer system 8MgO/Fe(001) at 1 keV (left side, Fig.~1c) and 6 keV
(right side, Fig.~1d).}
\end{figure}

\section{Treatment of correlation effects within dynamical mean field theory (DMFT)}
\label{dmft}
The following examples concerns the ferromagnetic transition metal systems Ni and Fe as prototype materials to study
electronic correlations and magnetism beyond the LSDA scheme. In Fig.~2 we present a comparison between experimental
photo emission data \cite{MBO+99} and calculated spectra using different theoretical approaches \cite{BME+06}. In the upper
row spin-integrated ARUPS measurements from Ni(011) along  ${\overline \Gamma}{\overline Y}$ for different angles of
emission are shown. The dotted lines represent the experimental data, whereas the solid lines denote the single-particle
approach to the measured spectral function. Obviously, the LSDA-based calculation completely fails to describe the 
experimental data. The energetic positions of the theoretical peaks deviate strongly from the measured ones. Furthermore,
the complicated intensity distributions that appear for higher angles of emission could not be reproduced by the
LSDA-based calculations at all. In contrast, the non self-consistent quasi-particle calculation provides a significant
improvement when compared to the measured spectra. For the complete variety of emission angles the energetic peak
positions coincidence with the experiment within about 0.1 eV. Only the overall shape of the measured spectral intensities
deviate from the calculations because of the neglection of multiple scattering and surface-related effects. In the
experiment the different peaks seem to be more broadened and the spectral weight especially for nearly normal emission
is shifted by about 0.1 eV to higher binding energies. In addition it seems that for very high emission angles like
60$^o$ an even more complicated peak structure is hidden due to limited experimental resolution. An additional
spin-analysis is therefore highly desirable for these experiments.
\label{figure2}
\begin{figure}[tp]
\includegraphics[width=0.45\textwidth,clip]{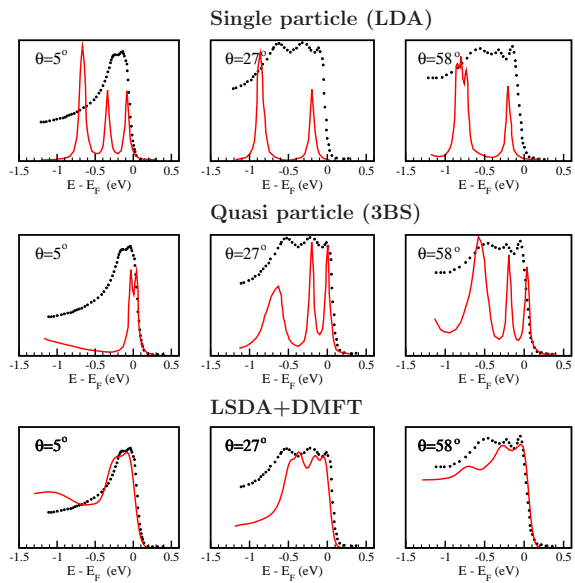}
\caption
{(color online) Spin-integrated ARUPS spectra from Ni(011) along  ${\overline \Gamma}{\overline Y}$ for three different 
angles of emission. Upper row: comparison between LSDA-based calculation and experiment \cite{MBO+99}; middle row: 
comparison between experiment and non self-consistent quasi-particle calculations neglecting matrix element and surface 
effects \cite{MBO+99}; lower row: spin-integrated LSDA+DMFT spectra including photo emission matrix elements (this
work). Theory: solid red line, experiment: black dots.} 
\end{figure}
\label{figure3}
\begin{figure}[tp]
\vspace*{-0.7cm}
\includegraphics[width=0.45\textwidth,clip]{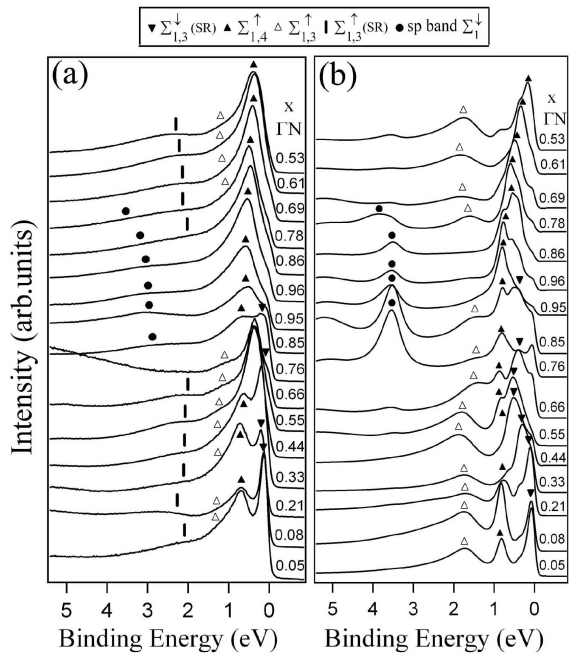}
\caption{(a) Experimental spin-integrated photo emission spectra of the Fe(110) surface measured with p-polarization in
normal emission along the $\Gamma$N direction of the bulk Brillouin zone. The curves are labeled by the wave vectors in
units of $\Gamma$N=1.55 $\AA^{-1}$. (b) Corresponding one-step model calculations based on the LSDA+DMFT method which 
include correlations, matrix elements and surface effects.}
\end{figure}

Within our work we could go far beyond previous theoretical studies by combining a sophisticated many-body approach as 
the self-consistent LSDA+DMFT method with the one-step based calculation of the corresponding spectral function.
The self-energy within the DMFT has been applied only for d-states and can be calculated in terms of two parameters - the
averaged screened Coulomb interaction $U$ and the exchange interaction $J$. The screening of the exchange interaction is
usually small and the value of $J$ can be calculated directly and is approximately equal to $0.9$~eV for all $3d$ elements.
This value has been adopted for all of our calculations presented here. For $U$ we used a value of $3$~eV. However, our test
calculations show that a somewhat different choice of $U$ does not substantially changes features and trends in the calculated
spectra. The intensity distributions resulting from the corresponding photo emission calculation are shown in the lower row
of Fig.~2. A first inspection reveals very satisfying quantitative agreement between experiment and theory for all emission
angles. Let us concentrate first on the excitation spectrum calculated for $\Theta=5^o$. The spin-integrated spectrum
exhibits a pronounced double-peak structure with binding energies of 0.1 eV and 0.3 eV. The second peak is slightly reduced
in intensity which is also in accordance with the experimental findings. Furthermore, the width of the spectral distribution
is quantitatively reproduced. The calculated binding energies are related to the real part of the self-energy that
corrects the peak positions due to dynamical renormalization procedure of the quasiparticles which is missing in a typical
LSDA-based calculations. The relative intensities and the widths of the different peaks, on the other hand, must be
attributed to the matrix-element effects which enter our calculations from the very beginning via the one-step model of
photo emission. The double-peak structure originates from excitation of the spin-split d-bands in combination with a
significant amount of surface-state emission \cite{EPHE80}. The two spectra calculated for high angles of emission show
the more broadened spectral distributions observable from the experimental data. An explanation can be given in terms of
matrix-element effects, due to the dominating dipole selection rules. The spin-resolved spectra reveal a variety of d-band
excitations in both spin channels, which in consequence lead to the complicated shape of the spectral distributions hardly
to be identified in the spin-integrated mode.

Our second example within this section concerns a spectroscopic study on ferromagnetic Fe \cite{SFB+09}. Figures 3a and 3b
display a comparison between spin-integrated ARPES data and theoretical LSDA+DMFT based one-step photo emission calculations
of Fe(110) along the $\Gamma$N direction of the bulk Brillouin zone (BZ) with p-polarized radiation. In our LSDA+DMFT
investigation underlying the ARPES calculations we use for the averaged on-site Coulomb interaction $U$ a value $U$=1.5
eV which lies between the experimental value $U\approx$1 eV \cite{SAS92} and a value $U\approx$2 eV derived from
theoretical studies \cite{CG05,CMK+08}. The $k$ values were calculated from the used photon energies
ranging from 25 to 100 eV. Near the $\Gamma$ point (k$\sim$0.06 $\Gamma$N), the intense peak close to the Fermi level
corresponds to a $\Sigma_{1,3}^{\downarrow}$ minority surface resonance, as indicated on top of Fig.~3. Experimentally,
its $\Sigma_{3}^{\downarrow}$ bulk component crosses the Fermi level at $k\sim$0.33 $\Gamma$N, leading to a reversal of
the measured spin-polarization and to a strong reduction of the intensity at $k=$0.68 $\Gamma$N in the minority channel.
The peak at the binding energy BE$\sim$0.7 eV, visible mainly for p-polarization in a large range of wave vectors between
$\Gamma$ and N, can be assigned to almost degenerate $\Sigma_{1,4}^{\uparrow}$ bulk-like majority states.
A $\Sigma_{3}^{\uparrow}$ feature at BE$\sim$1.1 eV dominates the spectrum at the $\Gamma$-point. Depending on the
polarization its degenerate $\Sigma_{1}^{\uparrow}$ states form a shoulder around the same BE. The broad feature
around 2.2 eV, visible at various $k$-points, but not at the N-point, is related to a majority $\Sigma_{1,3}^{\uparrow}$
surface state. Around the N-point (0.76$\leq k\leq $1.0) and at BE$\geq$3 eV we observe a $\Sigma_{1}^{\downarrow}$ band
having strong sp character. The pronounced difference between its theoretical and experimental intensity distributions
can be attributed to the fact that in the present calculations only local Coulomb repulsion between d electrons is
considered, without additional lifetime effects for the sp bands. Finally, we notice that the background intensity of the
spectrum at $k$=0.66 $\Gamma$N, corresponding to a photon energy of 55 eV, is strongly increasing by the appearance of
the Fe 3p resonance.

In conclusion, we have presented spectral function calculations of ferromagnetic Ni and Fe, which coherently combine an
improved description of electronic correlations with multiple-scattering, surface emission, dipole selection rules and
other matrix-element related effects that lead to a modification of the relative photo emission intensities. As has been
demonstrated, this approach allows on the one hand side a detailed and reliable interpretation of high-resolution angle-resolved
photo emission spectra of 3d-ferromagnets. On the other hand, it also allows for a very stringent test of new developments
in the field of DMFT and similar many-body techniques.

\section{Treatment of disordered alloys via the coherent potential approximation (CPA)}
\label{}
\label{figure4}
\begin{figure}[tp]
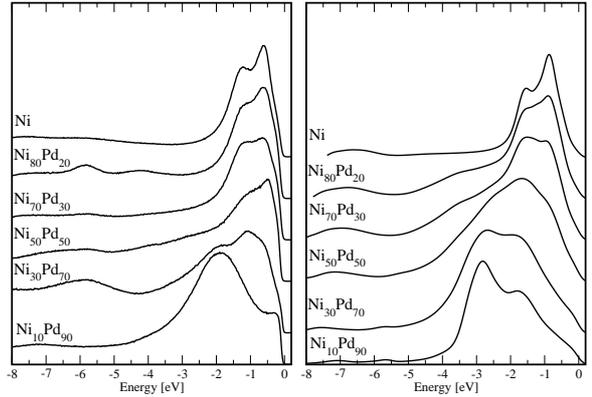

\includegraphics[width=0.23\textwidth,clip]{Fig4_A.eps}
\includegraphics[width=0.23\textwidth,clip]{Fig4_B.eps}
\caption {ARUPS spectra taken from the Ni$_{x}$Pd$_{1-x}$(001) alloy
surfaces as a function of the concentration x for a fixed photon
energy of $h\nu$=40.0 eV along $\Gamma$X in normal
emission. Experimental data shown in the left panel calculated spectra
presented in the right panel. Depending on the concentration $x$ a pronounced shift
in spectral weight towards the Fermi level is visible.}
\end{figure}
In this section we want to study the alloying effect in combination with electronic correlations. Fig.~4 shows a series
of spectra as a function of the concentration $x$ calculated for a photon energy $h\nu$=40 eV with linear polarized light.
The experimental data have been presented in the left panel and the corresponding LSDA+DMFT-based photo emission
calculations are presented in the right one. Our theoretical analysis shows that starting from the pure Ni case, the
agreement is fully quantitative with deviations less than 0.1 eV binding energy, as expected. Going to the Ni$_{0.80}$Pd$_{0.20}$
alloy the agreement is on the same level of accuracy concerning the range of binding energies between the Fermi energy
and 2 eV binding energy. Inspecting the density of states (DOS) for the Ni$_{0.80}$Pd$_{0.20}$ alloy this fact becomes
explainable, because this energy interval represents the Ni-dominated region. The Pd-states start to appear at about
2 eV below $E_{\rm F}$ besides of the small dip at the Fermi level. For higher binding energies the agreement is
also very good, although a bit more structure is observable in the theory, especially around 3.5 eV. An explanation for
this behavior can be found in terms of lifetime effects, but it should be mentioned here that the experimental intensity
background consisting of secondary electrons was not considered in the theoretical analysis. From the
Ni$_{0.70}$Pd$_{0.30}$ alloy system it becomes clearly visible that
the deviation between theory and experiment is mainly
introduced for increasing concentration of Pd. The same argument holds for the Ni$_{0.50}$Pd$_{0.50}$ and
Ni$_{0.30}$Pd$_{0.70}$ alloys shown next in the series. In addition, the spectra of Ni$_{0.30}$Pd$_{0.70}$ reveal some
deviations near the Fermi level. Also, the spectral intensity of the Ni surface resonance, that appears at about 0.5 eV
binding energy is underestimated in the calculation when compared to the experiment.

Our spectroscopical analysis has clearly demonstrated that the electronic properties of the Ni$_x$Pd$_{1-x}$ alloy
system depend very sensitively on the interplay of alloying and electronic correlation. A description within the LSDA
approach in combination with a CPA method results in a qualitative description of the electronic structure of
Ni$_x$Pd$_{1-x}$ only. 
\label{figure5}
\begin{figure}[tp]
\includegraphics[width=0.34\textwidth,angle=-90,clip]{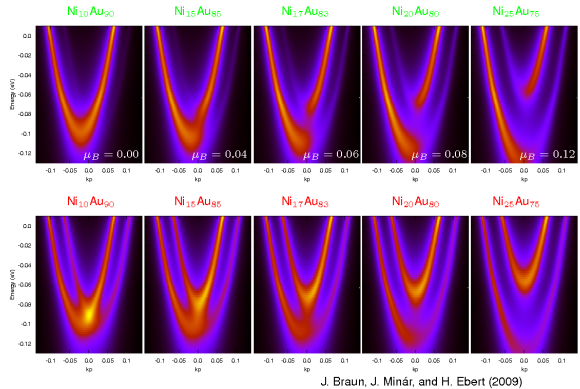}
\caption {ARUPS spectra taken from the Ni$_{x}$Au$_{1-x}$(111) alloy surfaces as a function of the concentration $x$
for a fixed photon energy of $h\nu$=21.0 eV along $\overline{\Gamma X}$. The upper panel shows the majority surface state,
the lower panel shows the minority surface state with increasing magnetic moment from left to right.} 
\end{figure}

The second example concerns more surface sensitive features, namely the spin-orbit-splitting observable for surface states
from high Z-materials like Au \cite{CDF+06}. Here we want to study the interplay between spin-orbit interaction and
correlation using as a prototype material the magnetic Au$_{x}$Ni$_{1-x}$ alloy. We are able to tune the magnetic moment
of the alloy as a function of the concentration $x$ and in consequence we can study both interactions on the same level
of accuracy. Fig.~5 shows a series of spin-resolved photo emission spectra from the Au$_{x}$Ni$_{1-x}$(111)-surface. For
a very low Ni concentration of about 5$\%$ the magnetic moment is nearly zero and we observe the pure spin-orbit split
Shockely surface state, well known from literature \cite{CDF+06}. With increasing the Ni concentration the magnetic moment
increases and the magnetic exchange interaction starts to compete with the spin-orbit splitting. Exchange causes a splitting
in energy, whereas spin-orbit interaction causes a splitting in k$_{\parallel}$. For Ni concentrations which lead to an
exchange splitting comparable with the spin-orbit one we find the interesting result that the two parabolas seem to shift
both in energy and k$_{\parallel}$ with the additional effect of structural distortion. For larger magnetic moments the
exchange interaction starts to win against the spin-orbit interaction and one can observe as usual the exchange-split
Shockely surface state which has to be expected on the Ni(111) surface.
 
These two examples may illustrate that the use of the CPA alloy theory self-consistently combined with a DMFT+LSDA approach 
serves as a powerful tool for electronic structure calculations, whereas the application of the fully relativistic one-step
model of photo emission, which takes into account chemical disorder and electronic correlation on equal footing guarantees
a quantitative analysis of the corresponding spectroscopical data.

\section{Effects of electron-phonon interaction in angle-resolved photo emission}
\label{phonons}

Nowadays high resolution photo emission measurements allow to investigate in very detail the electronic structure of
materials close to the Fermi level. This gives access to energy bands modified by electron-phonon
interaction, and therefore serves as a very important spectroscopical tool to characterize transport properties
or, for example, properties of super-conducting materials
\cite{CYKS00,RNE+00,TYK+01,REN+03}.
Modifications of the electronic structure generated by electron-phonon interaction are typically not accounted for
in LSDA-based band-structure calculations. The theoretical treatment of electron-phonon effects requires the use of
many-body techniques which originally had been developed to give a quantitative description of strong electronic
correlation effects. The self-energy approach often applied in many-body calculations can be mapped to the case of
electron-phonon related correlations. Therefore, having the electronic structure obtained by an ab-initio calculation,
the modifications due to electron-phonon interaction can be described via an electron-phonon self-energy
$\Sigma({\bf k},E_k)$ \cite{Gri81,HEC02} that appears in the momentum-resolved Green's function operator
\begin{eqnarray}
 G_k(z) &=& \frac{1}{z - E_k - \Sigma({\bf k},E_k)} \;.  
\label{Hspin_2}
\end{eqnarray}

%
%

The electron-phonon self-energy is derived using many-body perturbation theory within second order.
It follows \cite{Gri81}:
\begin{eqnarray}
\Sigma_k(E)& = & \int \Sigma_k(E,\omega) \alpha^2F_k(\omega) d\omega~, 
\end{eqnarray}
where $\Sigma_k(E,\omega)$ represents the self-energy obtained within the Einstein model \cite{Gri81}.
$\alpha^2F_k(\omega)$ denote the momentum-resolved Eliashberg function: 
\begin{eqnarray}
\alpha^2F_k(\omega) &\hspace*{-0.1cm}=\hspace*{0.0cm}& \nonumber
\hspace*{-0.3cm}  \sum_{\bf{q},\lambda,n'} 
|g_{\bf{k},\bf{k}+\bf{q}}^{\lambda,n,n'}|^2 \times \\
& & \hspace*{-1.0cm} \nonumber
\delta(E_{n,\bf{k}} - E_F)\delta(E_{n',\bf{k}-\bf{q}} - E_F)
\delta( \omega - \omega^{\lambda}_{\bf{q}}) \\
\end{eqnarray}
with the electron-phonon matrix element
\begin{eqnarray}
g_{\bf{k},\bf{k}+\bf{q}}^{\lambda,n, n'} & = &
\sum_{\nu,\alpha} 
\frac { \epsilon^{\bf{q},\lambda}_{\nu,\alpha}}
{ \sqrt{2M_{\nu} \omega^{\lambda}_{\bf{q}}}} 
\langle n,\bf{k} | \tilde{v}_{\bf{q}}^{\nu,\alpha} | n',\bf{k}-\bf{q}
\rangle~.
\end{eqnarray}
Herein, $\bf{k}, n$ stand for the wave vector and band index, respectively. $\bf{q}$ is the 
phononic wave vector and $\lambda$ denotes the number of phonon modes. $\epsilon^{\bf{q},\lambda}_{s,\alpha}$ defines
the polarisation of a phonon mode and $\nu$ numbers the atoms within
the unit cell.

To obtain explicitly the electron-phonon self-energy, we focus our attention on the Eliashberg function as the main
quantity determined through all coupled electronic and phononic states of the system. This function can be reformulated
in terms of the Green's function by use of the identity (see e.g. Butler (1985)). After some algebra the Eliashberg
function follows as the sum over all phonon states:
\begin{eqnarray}
\alpha^2F_k(\omega) 
& =& 
\sum_{{\bf q},\lambda} \sum_{\nu,\alpha} 
\frac { \epsilon^{{\bf q},\lambda}_{\nu,\alpha}\epsilon^{{\bf q},\lambda}_{\nu,\beta}}
{ 2M_{\nu} \omega^{\lambda}_{\bf{q}}} 
\delta( \omega - \omega^{\lambda}_{\bf{q}}) \\
&&\times \frac{1}{4} [T^{+,+}_{\alpha\beta,{\bf q}} + T^{-,-}_{\alpha\beta,{\bf q}} - 
T^{+,-}_{\alpha\beta,{\bf q}} - T^{-,+}_{\alpha\beta,{\bf q}})]~, \nonumber
\end{eqnarray}
with
\begin{eqnarray} \label{Eliashb_G2}
T_{\alpha\beta,{\bf q}}^{\pm,\pm} & \hspace*{-0.28cm} = \hspace*{-0.28cm} &
\int_{\Omega_0} d^3r d^3r' 
{\bf \nabla}V_{\nu,\alpha}({\bf r}) 
G_{{\bf k}-{\bf q}}^{\nu,\pm} 
{\bf \nabla}V_{\nu,\beta}({\bf r}')
G_{{\bf k}}^{\nu,\pm}.
\end{eqnarray}
According to Gy\"orffy and coworkers \cite{Wei90} the Green's function
$G_{\bf{k}}^{\nu,\pm}  = G_{\bf{k}}^{\nu}(r,r,E_{\rm F} \pm i0 )$
can be expressed in terms of the scattering path operator at the Fermi level 
$(\tau^{\nu}_{\bf{k}}(E_{\rm F} \pm i0))_{\Lambda'\Lambda}$ and is therefore accsessible in terms of multiple scattering theory.
 

\label{figure6}
\begin{figure}[tp]
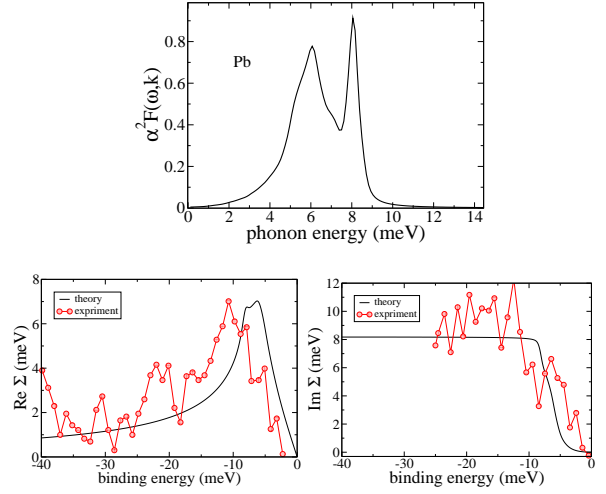

\begin{center}
\includegraphics[width=0.28\textwidth,clip]{Fig6_A.eps}\\
\end{center}
\includegraphics[width=0.23\textwidth,clip]{Fig6_B.eps}
\includegraphics[width=0.23\textwidth,clip]{Fig6_C.eps}
\caption {Upper panel: Calculated Eliashberg function
  $\alpha^2F_k(\omega)$ for Pb(110)  for
 ${\bf k} = \frac{\pi}{a}(0.17,1.,0.17)$. Lower panel: corresponding
k-dependent self-energy, real part shown on left, imaginary part shown on the right.}
\end{figure}
The complex self-energy contributes due to electron-electron and electron-impurity scattering,
as well as due to electron-phonon scattering described by the electron-phonon self-energy
$\Sigma_{el-ph}$. The real and imaginary parts of $\Sigma_{el-ph}$ are responsible for band renormalisations
around $E_{\rm F}$. These quantities are determined by electron-phonon scattering induced quasiparticle lifetimes,
measurable in photo emission experiments. To demonstrate the effect of electron-phonon interactions on the electronic
quasi-particle states near $E_{\rm F}$ we present photo emission calculations for Pb, which serves as a prototype system
and compare our theoretical results to corresponding experimental data \cite{CDF+06}. Pb is of special interest
because of the presence of strong electron-phonon interactions which in particular lead to the superconducting
state at low temperatures \cite{CYKS00,REN+03}.

The calculations have been performed for energies close to the Fermi level. The wave vectors in the BZ
(${\bf k} = \frac{\pi}{a}(0.17,1.,0.17)$) correspond to the experimental geometry \cite{REN+03}. Fig.~\ref{figure6}
shows a comparison between the experimental and theoretical real and
imaginary parts of the self-energy $\Sigma_k$.  Obviously, the
agreement is very satisfying.

\label{figure7}
\begin{figure}[t]
\includegraphics[width=0.45\textwidth,clip]{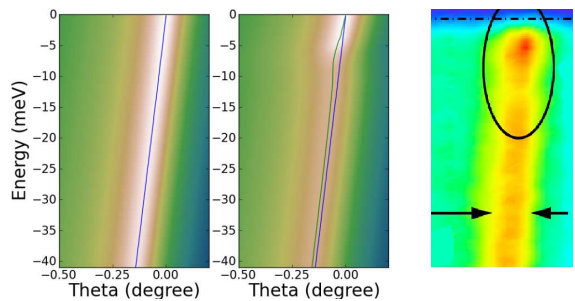}
\caption {ARUPS spectra taken from the Pb(110) surface 
excluding (left panel) and including (middle padel) phonons calculated self-consistently from first
principles. Corresponding experimental data are shown on the right panel \cite{ELT05}.}
\end{figure}
In general, it is possible to deduce a phononic self-energy from measured photo emission spectra. The
imaginary part of the self-energy can be estimated from the full width at half maximum of the energy distribution curves
fitted by a Lorenzian. The real part is than obtained by a Kramers-Kronnig transformation. However, in cases where
for example several transitions appear close to the Fermi level or even additional surface contributions to the photocurrent
take place, this analysis might be questionable. Therefore, it seems to be quite reasonable to make a direct
comparison between angle resolved photo emission calculations and corresponding experimental data. For this purpose
we developed a scheme that allows to include directly a ${\bf k}$-dependent self-energy into the one-step model
of photo emission. In Fig.~7 we present photo emission spectra calculated for the two cases without (left pannel) and
with (middle panel) electron-phonon interactions and compare them to the correponding measurements \cite{ELT05}.
The contour plot has been calculated for HeI light along $\bar{\Gamma}\bar{X}$. More precisely, bands forming the
locally tubular Fermi surface of Pb(110) crosses E$_F$ at about $5^o$ off-normal emission along $\bar{\Gamma}\bar{X}$. 
In the middle pannel of Fig.~7 a clear renormalisation of the bare band (often called {\it kink}) at $8$meV binding
energy is observable. This effect appears just where the real part of the self-energy reaches its maximum. Futhermore,
an additional broadening of the spectral features due to the imaginary part of the phononic self-energy is visible
in the calculated spectra. This, of course, completes a lifetime analyses of spectral features in a quantitative
sense because the imaginary part of the phononic self-energy can be identified as the last contribution that has to
be considered in addition to impurity scattering and electronic correlation related broadening mechanisms of
photo emission intensities. 
\section{Conclusions}
\label{conclusions}
We have shown several examples of photo emission calculations which reflect the wide range of applicability of the
one-step model of photo emission. In its latest version that is directly combined with the SPRKKR program package
quantitative spectroscopical investigations may range from 10 eV up to 10 keV concerning the photon energy. This
gives not only direct accsess to the UV regime but also to the so called HAXPES-regime of angle resolved photo emission.
Due to the LSDA+DMFT method the range of applicability has been greatly enhanced ranging now from simple paramagnetic
metals to complex layered compounds inhibiting strong electronic correlations. The recent extent of the one-step model
to disorderd magnetic alloys round up the range of applicability to a variety of complex 3D- as well as 2D-materials
and allows investigations on surface related features for low photon energies as well as bulk like analyses for higher
photon energies. As it holds for photocurrent calculations in the HAXPES regime phonon related effects are also very
important when analysing the intensity distributions which are obtained for energies around the Fermi level. The second
point has been addressed in the last section in which we demonstrated how to introduce in a quatitative sense a k-dependent
phononic self-energy that allows for spectroscopical calculations from kink-like structures. Therefore this ansatz enlarges
in combination with the LSDA+DMFT method again the range of applicability to systems like high T$_c$ cuprats or other highly
correlated classes of materials.

\section{Acknowledgments}

Financial support by the Deutsche Forschungsgemeinschaft through FOR
1346, EB-154/18,EB-154/23  and MI-1327/1  is gratefully acknowledged. We would
like to thank A.~Eiguren and C.~Ambrosch-Draxl for giving access to
the program calculating electron-phonon self-energy.








\end{document}